\documentclass{ws-ijmpc}
\usepackage{graphicx}
\usepackage{lipsum}
\usepackage{enumerate}
\usepackage{amsmath}
\usepackage{amssymb}
\usepackage{color}
\usepackage{bm}
\usepackage{enumitem}
\setenumerate{fullwidth,itemindent=
\parindent,listparindent=\parindent,itemsep=0ex,partopsep=0pt,parsep=0ex}
\usepackage[square, comma, sort&compress, numbers]{natbib}
\usepackage[colorlinks,urlcolor=blue,linkcolor=blue,anchorcolor=blue,citecolor=blue]{hyperref}
\begin{document} 

\title{Occurrence of synchronized flow due to overtaking strategy in the Nagel-Schreckenberg model}
\author{Zhu Su, Weibing Deng, Jihui Han, Wei Li and Xu Cai\footnote{Email:
wdeng@mail.ccnu.edu.cn, liw@mail.ccnu.edu.cn}}
\address{Complexity Science Center, Institute of Particle
Physics,\\ Hua-Zhong (Central China) Normal University, Wuhan
430079, China}
\maketitle
\begin{abstract} 
The Nagel-Schreckenberg model with overtaking strategy (NSOS) is proposed, and numerical simulations are performed for both closed and open boundary conditions. The fundamental diagram, space-time diagram, and spatial-temporal distribution of speed are investigated. In order to identify the synchronized flow state, both the correlation functions (autocorrelation and cross-correlation) and the one-minute average flow rate vs. density diagram are studied. All the results verify that synchronized flow does occur in our model. 
\end{abstract}
\keywords{Traffic model; synchronized flow; overtaking strategy.}
\section{Introduction}
 With the development of urbanization, traffic problems have become an urgent matter. For example, vehicular traffic has rapidly outstripped the capacities of the nation's highways. It is increasingly necessary to understand the dynamics of traffic flow. Since the behavior of traffic flow is not governed by the general principles of thermodynamics and statistical mechanics \cite{Helbing2001}, different models of traffic flow have been proposed \cite{Chowdhury2000a}, such as car-following models \cite{Pipes1953,Zhang2005,Tang2015} which are based on the basic principles of classical Newtonian dynamics, and the cellular automaton (CA) \cite{Wolfram1983} models which describe the traffic in terms of the stochastic dynamics of individual vehicles. CA-based microscopic traffic models are promising tools for large-scale computer simulations due to their simplicity. One of the CA traffic models is the one-dimensional NS model \cite{Nagel1992}, which is considered to be the simplest one to study the traffic flow, and is able to reproduce some basic phenomena of real traffic. Based on the NS model, some modified models were developed by imposing more conditions to make it realistic, such as the Velocity Effect (VE) model \cite{Li2001}, the Comfortable Driving (CD) model \cite{Knospe2000}, and the relative multi-lane models \cite{Jia2004,Hu2012,Li2006,Feng2015,Lv2011}.\\
 \indent However, the NS model belongs to the framework of two-phase traffic theory \cite{Helbing2001,Chowdhury2000a}, in which a transition from free flow to wide moving jams ($F \rightarrow J$) happens if perturbed. Based on the empirical data, Kerner \cite{Kerner1997,Kerner1998,Kerner2009,Kerner2011,Kerner2013,Kerner2014} proposed a three-phase traffic theory, which considered the free flow, the synchronized flow and the wide moving jams. According to the three-phase theory, the wide moving jams do not emerge spontaneously in free flow. Instead, there is a sequence of two first-order phase transitions: firstly the transition from free flow to synchronized flow occurs ($F \rightarrow S$), then at different locations the wide moving jams emerge in the synchronized flow ($S \rightarrow J$). \\
 \indent Recently, some new microscopic models \cite{Jiang2005,Wang2007,Tian2009,Larraga2010} based on the three phase traffic theory have been proposed, which can display the synchronized pattern features. For example, Jiang \cite{Jiang2003} presented a CA model based on the Knospe's comfortable driving model \cite{Knospe2000}, which can reproduce the synchronized flow. Moreover, Lee \cite{Lee2004} proposed a different CA model which would also describe the features of synchronized flow.\\
\begin{figure}[ht]
	\centering
	\includegraphics[scale=0.3]{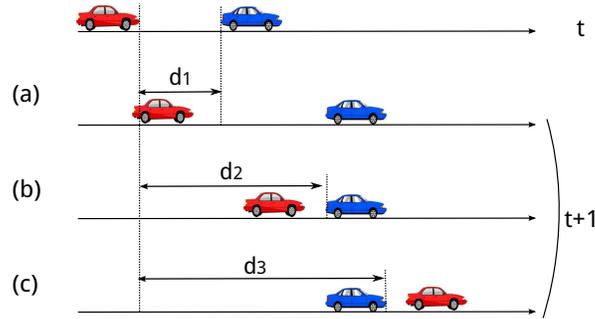}
	\caption{\label{graph}(Color online) Schematic of a real traffic situation, which shows an important mechanism of our model that was not considered in the original NS model.}
\end{figure}
\indent In this paper, we introduced a modified NS model which can also reproduce the synchronized flow. In real traffic, some vehicles would try to overtake the preceding ones, no matter whether they can succeed or not. This behavior leads to two effects: (i) The vehicles behind will follow the preceding ones, and move into the regime where their preceding ones will go through in the next step. (ii) The vehicles behind will overtake the preceding ones successfully. By considering these facts, our paper proposes overtaking strategy (OS) based on the NS model to describe the overtaking phenomenon (thereafter abbreviated as ``NSOS model"). Fig.\,\ref{graph} (a) shows the mechanism of the NS model, the red car is only able to move within the gap $d_1$ at time $t+1$. Unlike the vehicles in the NS model, overtaking vehicles in the NSOS model have much more space to move. For example, as shown in Fig\,\ref{graph} (b), even though the red overtaking car can not overtake the blue car, it could move much further within the gap $d_2$. While, if it overtakes successfully, it could move longer than the distance $d_3$ shown in Fig\,\ref{graph} (c). \\
\indent The paper is organized as follows: in Section $2$, we introduced our model. In Section $3$, we discuss the simulation results, including the fundamental diagram, the space-time diagram and the spatial-temporal distribution of speed. Furthermore, the correlation function analysis is also given. Simulation results are obtained for both closed and open boundary conditions. Summary is given in the last section.
	
\section{The NSOS model}
Based on the NS model, we propose an overtaking strategy to explore its effect on the traffic flow on the one-dimensional lattices. (1) We pick up overtaking vehicles with probability $q$ at each time step. This means that every vehicle could be an overtaking one with probability $q$ at each time step, and every time step the system has $qN$ overtaking vehicles on average. Overtaking vehicle will try to overtake the preceding one, but not all could succeed, which depends on the preceding one's configuration at next time step. (2) To avoid collisions, the overtaking vehicle decelerates if it reaches the preceding one's position. (3) For simplicity, each overtaking vehicle is only able to overtake one vehicle each time, and its position will locate in front of its preceding vehicle if it overtakes successfully. (4) If two consecutive vehicles are both overtaking ones, and the preceding one overtakes successfully, then the one behind is not able to overtake successfully any more. (5) The velocities of overtaking vehicles are decreased by one with probability $p$ except for successfully overtaking vehicles. (6) If the vehicle is not an overtaking one (here we name it as an ordinary vehicle), its velocity will be updated according to the rules of the NS model. The detailed updating rules are as follows:
\begin{description}
	\item[1.] Pick up the overtaking vehicles with probability $q$.
	\item[2.] Update the velocity:
	\begin{description}
	\item[(I)] For the ordinary vehicles:
	\begin{description}
	\item[(1)] Acceleration: \\
	$v(j,t_{1}) \rightarrow min(v(j,t)+1, v_{max})$.
	\item[(2)] Deceleration: \\
	$v(j,t_{2}) \rightarrow min(v(j,t_{1}),d(j,t))$.
	\item[(3)] Random braking: \\
	$v(j,t_3) \rightarrow max(v(j,t_{2})-1,0)$ with the probability $p$.
	\end{description}
	\item[(II)] For the overtaking vehicles:
	\begin{description}
	\item[(1)] Acceleration: \\
	$v(j,t_{1}) \rightarrow 	min(v(j,t)+1, v_{max})$.
	\item[(2)] If $v(j,t_{1}) > d(j,t) +v(j+1,t+1)$, the position $d(j,t)+v(j+1,t+1)+l$ is empty and the $(j+1)$th vehicle does not overtake successfully,
		\item[(i)] Overtaking: \\
		$v(j,t_{3}) \rightarrow d(j,t)+v(j+1,t+1)+l.$
	\item[(3)] Otherwise,
		\item[(i)] Deceleration: \\
		$v(j,t_{2}) \rightarrow min(d(j,t)+v(j+1,t+1)-al,v(j,t_{1}))$.
		\item[(ii)] Random braking with probability $p$: \\
		$v(j,t_{3}) \rightarrow min(v(j,t_{2})-1,0)$.
	\end{description}	
	\end{description}		
	\item[3.] Movement:\\
	 $x(j,t+1)=x(j,t)+v(j,t_{3})$.
\end{description}
\indent Here, $v(j,t)$ denotes the velocity of the $j$th vehicle at time $t$, $x(j,t)$ denotes its corresponding position and $l$ is the length of a single vehicle. The number of empty sites in front of the $j$th vehicle is denoted by $d(j,t)=x(j+1,t)-x(j,t)-l$. To avoid collision, we assume $a=2$ if the $(j+1)$th vehicle is also the overtaking one and overtakes successfully. In other cases, $a$ equals one.\\
\indent In the following section, simulations are performed under both closed and open boundary conditions. Since the velocity of an overtaking vehicle at time $t+1$ is relative to its preceding vehicle's velocity, we need to know the preceding vehicle's velocity at time $t+1$ first. Since the velocities of ordinary vehicles are independent of the preceding ones', we could pick up an ordinary vehicle as the leading one and update its velocity first, and then update its rear vehicle's velocity. In our simulations, we choose the first vehicle and the last one as ordinary vehicles all the time and update their velocities first, and then update others' velocities.
\section{Simulation results}
\indent In the simulations, the system size is taken to be $L=10000$. The length of a lattice corresponds to $1.5$ m, a vehicle occupies five lattices and one time step corresponds to $1$ s. It is assumed that $v_{max}$ equals $25$. Initially, $N$ vehicles are randomly distributed on the lattices and the velocity of each vehicle is designated by an integer randomly chosen from zero to $v_{max}$. One simulation runs for $20000$ time steps. The sampling are collected when the time evolution reaches to the $10000$th time step, and each data point is averaged over $100$ different initial configurations.
\begin{figure}[ht]
	\centering
	\includegraphics[scale=0.25]{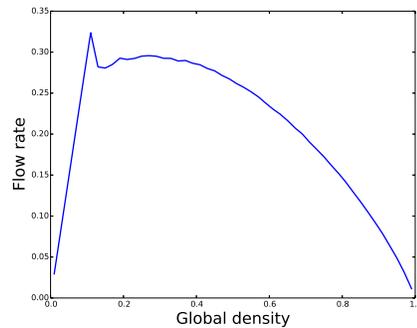}
	\caption{\label{fundamental}(Color online) The fundamental diagram of NSOS model obtained on a circular road in the case of $p=0.25$ and $q=0.5$. The synchronized flow state occurrs in the plateau regime.}
\end{figure}
\subsection{Periodic boundary conditions}
In this section, we firstly show the simulation results on a closed road with periodic boundary conditions. We plot some simulation results, such as the fundamental diagram, the space-time diagram and the spatial-temporal distribution of speeds. In order to demonstrate the existence of synchronized flow further, the autocorrelation and cross correlation functions are studied.
\begin{figure}[ht]
	\centering
	\includegraphics[scale=0.35]{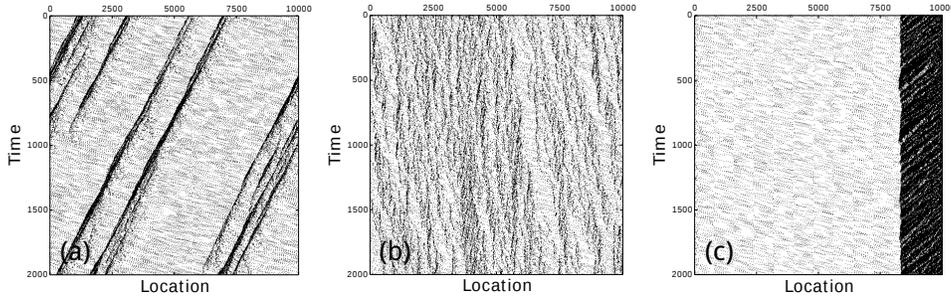}
	\caption{\label{spacetime1}(Color online) The space-time diagram of our model in the case of $p=0.25$ and $\rho=0.3$. (a) The wide moving jams phase when $q=0$ (NS model). (b) The synchronized flow phase when $q=0.5$. (c) The phases separation when $q=1$.}
\end{figure}

\begin{figure*}[htbp]	
		\centering
		\includegraphics[scale=0.33]{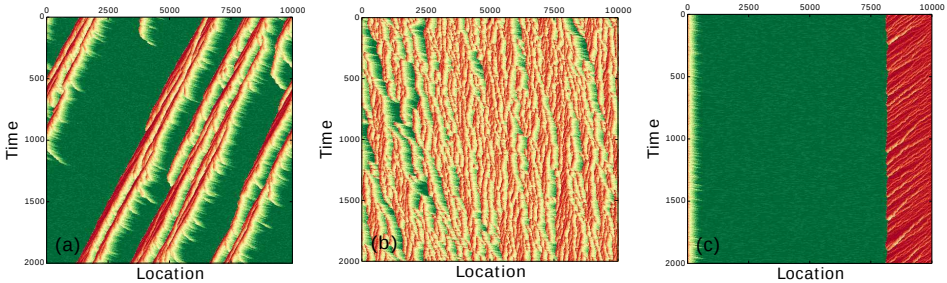}
	\caption{\label{speedtime}(Color online) The spatial-temporal distributions of speed for the different overtaking probabilities $q$ in the case of $p=0.25$ and $\rho=0.3$ under the closed boundary condition. (a) Only wide moving jams exist in the congested state when $q=0$. (b) Synchronized flow could be observed when $q=0.5$. (c) When $q=1$, free flow and congested phases are evidently separated, which is called phases separations. The green region is related to the state of free flow with the speed $v$ approaching the maximum velocity, the yellow region is related to the state of synchronized flow and the red region is related to the state of wide moving jams.}
\end{figure*}
\indent In order to study the influence of overtaking probability $q$ on the traffic flow, we plot the fundamental diagram in Fig.\,\ref{fundamental}. One would see that: (i) In the low-density region, there is just free flow, and all the vehicles move with $v_{max}$. (ii) In the second region, as the density increases, the flow rate is almost the maximum flow and does not decline, which is the typical feature of synchronized flow. (iii) In the high density region, the flow rate decreases with the increasing of density. \\
\indent Fig.\,\ref{spacetime1} presents the snapshots of space-time diagrams of the NSOS model with different values of $q$. The space direction is horizontal, the time coordinate is downward. When $q=0$, our model becomes the NS model, the congestion state only exits wide moving jams, as shown in Fig.\,\ref{spacetime1} (a). When $q>0$, such as $q=0.5$, the road is occupied by both overtaking vehicles and ordinary vehicles. As shown in Fig.\,\ref{spacetime1} (b), we find congestion clusters are very different from those of the NS model. Unlike the NS model, in which the congestion clusters move upstream, the clusters in the NSOS model are fixed at the same position, which is the feature of synchronized flow. When $q=1$, all the vehicles are overtaking vehicles if we ignore the boundary condition (we assigned the first and last vehicle as the ordinary vehicles). From the space-time diagram shown in Fig.\,\ref{spacetime1} (c), we can see that the free flow and congested phases are evidently separated, which is called phase separations. In fact, this interesting phenomenon is induced by the boundary condition. \\
\indent Next, we present the spatial-temporal distributions of speed for different overtaking probabilities $q$. As shown in Fig.\,\ref{speedtime} (a), when the overtaking probability $q=0$, the model returns to the original NS model, there only exist the free flow and wide moving jams. When $q=0.5$, we can see the synchronized flow dominates in all the regime. Again, when $q=1$, shown in Fig.\,\ref{speedtime} (c), we can see the free flow and congested phases are evidently separated.\\
\begin{figure}[ht]
	\centering
	\includegraphics[scale=0.3]{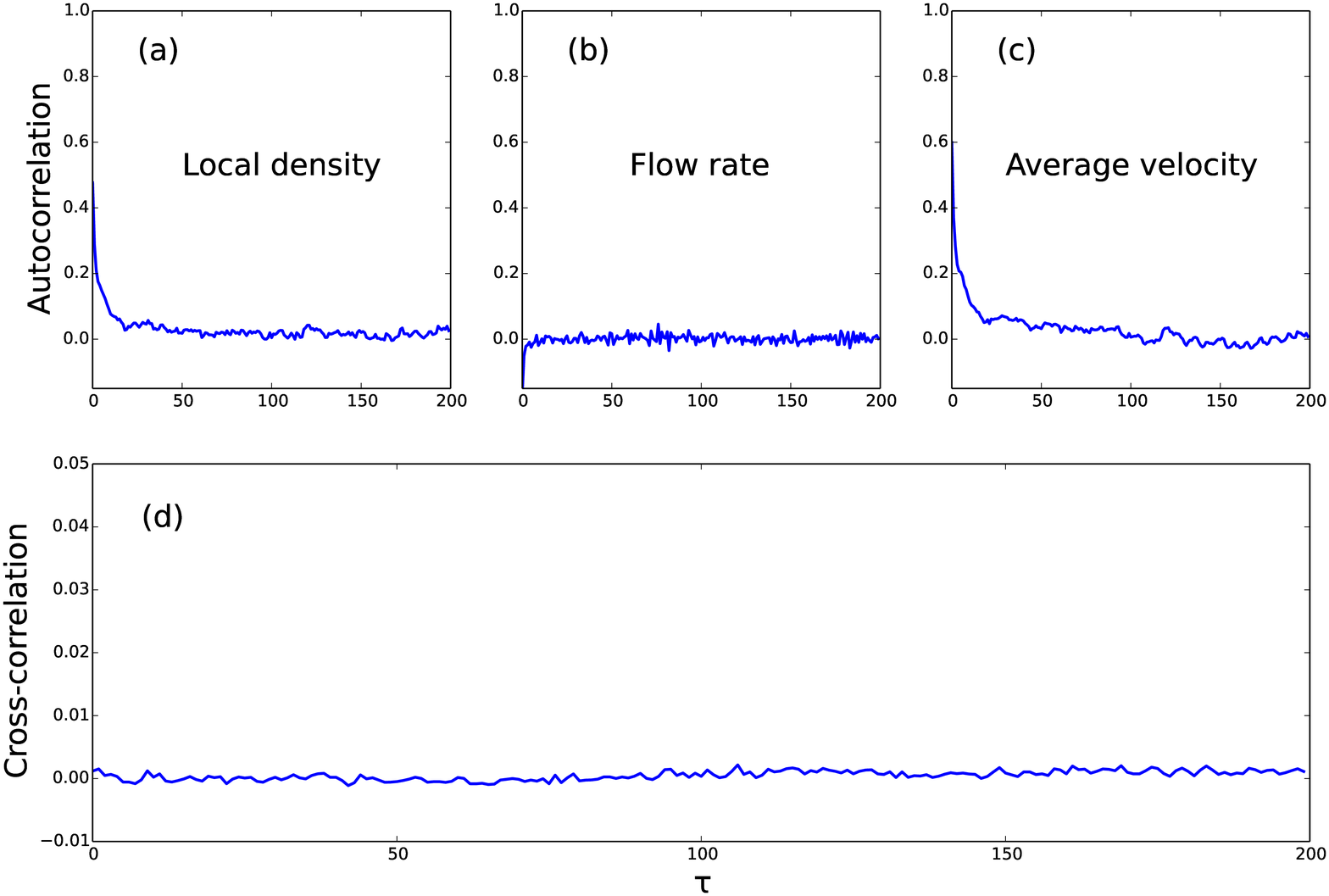}
	\caption{\label{correlation}(Color online) (a)-(c) The autocorrelation functions of one-minute aggregate of local density, flow rate and average velocity of synchronized flow, respectively. (d) Cross-correlation function between density and flow rate of the synchronized flow. The global density is $\rho=0.3$. }
\end{figure}
\begin{figure}[ht]
	\centering
	\includegraphics[scale=0.35]{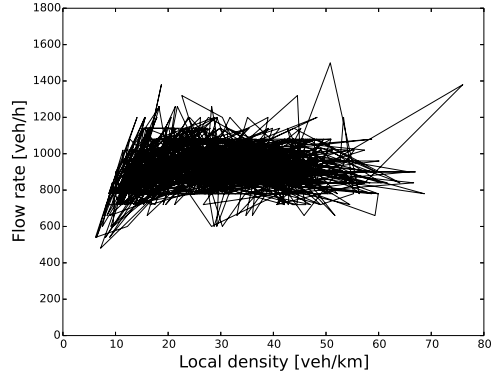}
	\caption{\label{flux}(Color online) The one minute averaged flow rate vs. density diagram corresponding to the synchronized flow in Fig.\,\ref{correlation}.}
\end{figure}
\indent In order to identify the synchronized flow more clearly, the correlation function is investigated. Time series data are obtained through a fixed virtual loop detector. First we analyze the autocorrelation of the aggregated quantities $x(t)$ \cite{Neubert1999} $$a_x(\tau)=\frac{\left\langle x(t)x(t+\tau)\right\rangle-\left\langle x(t)\right\rangle ^2 }{\left\langle x^2(t)\right\rangle -\left\langle x(t)\right\rangle ^2}$$, where the brackets $\left\langle \dddot{} \right\rangle $ indicate the average over all the series of $x$. In Figs.\,\ref{correlation} (a)-(c), the autocorrelations of one-minute aggregates of the density, flow flux and average velocity of a synchronized flow are shown. One could find the autocorrelations are all close to zero, which means no long-range correlations exist. Furthermore, Fig.\,\ref{correlation} (d) shows that the cross-correlation function between density and flow flux vanishes in large time scale \cite{Neubert1999} $$ c_{xy}(\tau)=\frac{\left\langle x(t)y(t+\tau)\right\rangle -\left\langle x(t)\right\rangle \left\langle y(t)\right\rangle }{\sqrt{\left\langle x(t)^2\right\rangle -\left\langle x(t)\right\rangle ^2}\sqrt{\left\langle y(t)^2\right\rangle -\left\langle y(t)\right\rangle ^2}}.$$ According to Ref.\,\cite{Neubert1999}, ``neither the free flow nor the jams has zero-valued correlation functions characteristically". Both correlation functions exhibit characteristics of synchronized flow. Fig.\,\ref{flux} shows the one-minute averaged flow rate vs. density diagram covering a two-dimensional region in the plane, which is consistent with the fundamental hypothesis of three-phase traffic theory \cite{Kerner1997}.\\
\indent Finally, we try to explore the reason why the NSOS model could reproduce the synchronized flow. According to Kerner's three-phase theory \cite{Kerner2013,Kerner2014}, the characteristic feature of the three-phase theory assumes the existence of two qualitatively different instabilities in vehicular traffic: (1) The instability associated with the over-acceleration, causing a growing wave of vehicle speed increase. (2) The instability of the GM model class associated with the over-deceleration effect that causes a growing wave of speed reduction. In the NSOS model, which includes the over-deceleration effect coming from the NS model, the overtaking mechanism provides the over-acceleration effect, which reproduces the synchronized flow. In detail, if a vehicle is an overtaking one with probability $q$, it will have more space to move, such as the case in Fig.\,\ref{graph} (b)-(c), this gives the overtaking vehicle the ability of over-acceleration. In fact, the gap $d_2$ in Fig.\,\ref{graph} (b) is similar to the effect distance in \cite{Knospe2000,Jiang2005}, which also considers the information of preceding vehicle including its position and velocity and reproduces the synchronized flow. We believe that the further distance that the overtaking vehicle could move is the reason that the synchronized flow occurs in our model.
\subsection{Open boundary conditions}
\indent Next we show the simulation results induced by an on-ramp under open boundary conditions. For open boundary conditions, vehicles enter a road from the left end of the road and move out of the road from the right end. We assume that the left-most lattice on the road corresponds to $x=1$ and the position of the left-most vehicle is $x_{l}$. If $x_{l}>v_{max}$, a new vehicle with velocity $v_{max}$ will be injected to the position $min\left\lbrace x_{l}-v_{max},v_{max} \right\rbrace$ with probability $q_{in}$. At the right boundary, the leading vehicle moves without any limits. When the position of the leading vehicle is beyond $L$, it will be removed and its following vehicle becomes the leader.\\
\begin{figure}[ht]
	\centering
	\includegraphics[scale=0.29]{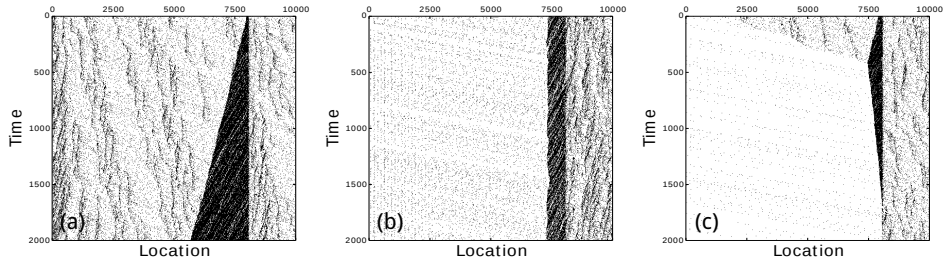}
	\caption{\label{spacetimeopen}(Color online) The space-time diagrams of the NSOS model for $v_{max}=25$, $q=0.3$ and $p=0.25$ on an open road. (a) WSP: Widening synchronized flow pattern $q_{in}=0.4$, $q_{on}=0.5$. (b) LSP: Localized synchronized flow pattern $q_{in}=0.2$, $q_{on}=0.2$ (c) DGP: Dissolving general pattern. $q_{in}=0.04$, $q_{on}=0.8$. }
\end{figure}
\indent At the on-ramp, we scan the region of the on-ramp $\left[ x_{on}-L_{ramp}, x_{on}\right]$ and find out the longest gap in this region. If this gap is long enough for a vehicle, a new vehicle will be injected into the lattice in the middle of the gap with probability $q_{on}$. We set its velocity as the velocity of its preceding one. In our simulations, we assign $L=10000$, $x_{on}=0.8L$ and $L_{ramp}=30$.\\
\indent If the parameters $q_{in}$ and $q_{on}$ are fine-tuned, the NSOS model could simulate different congested patterns, which are predicted by three-phase traffic theory. Fig.\,\ref{spacetimeopen} is the space-time diagrams in various values of $q_{in}$ and $q_{on}$. As Fig.\,\ref{spacetimeopen} (a) shows, the parameters are $q_{in}=0.4$ and $q_{on}=0.5$, at the on-ramp, we can see the upstream front is continuously widening upstream. This pattern is named the ``widening synchronized flow pattern" (WSP). Fig.\,\ref{spacetimeopen} (b) shows another congested pattern, which in the case of $q_{in}=0.2$ and $q_{on}=0.2$, the downstream front of the synchronized flow is fixed at the on-ramp, in contrast to the WSP, the upstream front of this synchronized flow is not continuously widening over time, but limited somewhere upstream of the on-ramp. The whole synchronized region is localized near the on-ramp, therefore, this pattern is called the ``localized synchronized flow pattern" (LSP). When $q_{in}=0.04$ and $q_{on}=0.8$, shown in Fig.\,\ref{spacetimeopen} (c), the transition from synchronized flow to jams occurs inside the WSP, but it could not induce wide moving jams sequences, but only a jamming area dissolving over time, which is called ``dissolving general pattern" (DGP). \\
\indent The patterns in the Fig.\,\ref{spacetimeopen} are consistent with the three-phase theory, and synchronized flow does occur in this model.
\section{Conclusions}
\indent The NSOS model is an extension of the NS model with overtaking strategy. The simulation results indicate that this model can reproduce the complicated traffic behavior of real traffic, such as the phenomena of synchronized flow and phase separations. This model restores to the original NS model by assigning the overtaking probability $q=0$.\\
\indent In this paper, we have analyzed the fundamental diagram and shown the space-time diagrams of different phases. The autocorrelation and cross-correlation functions of the synchronized flow state are investigated. The one-minute average flow rate vs. density diagram is presented and it is satisfactorily consistent with the empirical data. The underlying mechanism has been analyzed: The overtaking strategy provides equivalent effects as the ``over-acceleration noise", and the further distance that the overtaking vehicle could move is the main reason that synchronized flow occurs in our model. Furthermore, on the open boundary condition with an on-ramp, this model could reproduce synchronized flow and various congested patterns.\\
\section*{Acknowledgments}

This work was supported by National Natural Science Foundation of China (Grant No. 11505071), the Programme of Introducing Talents of Discipline to Universities under Grant No. B08033 and the Fundamental Research Funds for the Central Universities (Grant No. KJ02072016-0170, CCNU).


\begin{thebibliography}{99}

\bibitem{Helbing2001} D. Helbing, {\it Rev. Mod. Phys.} {\bf 73} 1067 (2001).
\bibitem{Chowdhury2000a} D. Chowdhury, L. Santen, and A. Schadschneider, {\it Phys. Rep.} {\bf 329} 199 (2000).
\bibitem{Pipes1953} L. A. Pipes, {\it J. Appl. Phys.} {\bf 24} 274 (1953).
\bibitem{Zhang2005} H. M. Zhang and T. Kim, {\it Transp. Res. B} {\bf 39} 385 (2005).
\bibitem{Tang2015} T.Q. Tang, J. G. Li, S. C. Yang and H.Y. Shang, {\it Physica A} {\bf 419} 293 (2015).
\bibitem{Wolfram1983} S. Wolfram, {\it Rev. Mod. Phys.} {\bf 55} 601 (1983).
\bibitem{Nagel1992} K. Nagel and M. Schreckenberg, {\it J. Phys. I France} {\bf 2} 2221 (1992).
\bibitem{Li2001} X. Li, Q. Wu and R. Jiang, {\it Phys. Rev. E} {\bf 64} 066128 (2001).
\bibitem{Knospe2000} W. Knospe, L. Santen, A. Schadschneider and M. Schreckenberg, {\it J. Phys. A: Math. Gen.} {\bf 33} L447 (2000).
\bibitem{Jia2004} B. Jia, R. Jiang and Q. S. Wu, {\it Int. J. Mod. Phys. C} {\bf 15} 381 (2004).
\bibitem{Hu2012} X. Hu, W. Wang and H. Yang, {\it Physica A} {\bf 391} 5102 (2012).
\bibitem{Li2006} X. G. Li, B. Jia, Z. Y. Gao and R. Jiang, {\it Physica A} {\bf 367} 479 (2006).
\bibitem{Feng2015} S. Feng, J. Li, N. Ding and C. Nie, {\it Physica A} {\bf 428} 90 (2015).
\bibitem{Lv2011} W. Lv, W. G. Song and Z. M. Fang, {\it Physica A} {\bf 390} 2303 (2011).
\bibitem{Kerner1997} B. S. Kerner, {\it Phys. Rev. Lett.} {\bf 79} 4030 (1997).
\bibitem{Kerner1998} B. S. Kerner, {\it Phys. Rev. Lett.} {\bf 81} 3797 (1998).
\bibitem{Kerner2009} B. S. Kerner and S. L. Klenov, {\it Phys. Rev. E} {\bf 80} 056101 (2009).
\bibitem{Kerner2011} B. S. Kerner, S. L. Klenov and M. Schreckenberg, {\it Phys. Rev. E} {\bf 84} 046110 (2011).
\bibitem{Kerner2013} B. S. Kerner, S. L. Klenov, G. Hermanns and M. Schreckenberg, {\it Physica A} {\bf 392} 4083 (2013).
\bibitem{Kerner2014} B. S. Kerner, S. L. Klenov and M. Schreckenberg, {\it Phys. Rev. E} {\bf 89} 052807 (2014).
\bibitem{Jiang2005} R. Jiang and Q. S. Wu, {\it Eur. Phys. J. B} {\bf 46} 581 (2005).
\bibitem{Wang2007} R. Wang, R. Jiang, Q. Wu and M. Liu, {\it Physica A} {\bf 378} 475 (2007).
\bibitem{Tian2009} J. Tian, B, Jia, X. Li, R. Jiang, X. Zhao and Z. Gao, {\it Physica A} {\bf 23} 4827 (2009).
\bibitem{Larraga2010} M. E. Lárraga and L. Alvarez-Icaza,  {\it Physica A} {\bf 389} 5425 (2010).
\bibitem{Jiang2003} R. Jiang and Q. S. Wu, {\it J. Phys. A: Math. Gen.} {\bf 36} 381 (2003).
\bibitem{Lee2004} H. K. Lee, R. Barlovic, M. Schreckenberg and D. Kim, {\it Phys. Rev. Lett.} {\bf 92} 23 (2004).
\bibitem{Neubert1999} L. Neubert, L. Santen, A. Schadschneider and M. Schreckenberg, {\it Phys. Rev. E} {\bf 60} 6480 (1999).
\end{thebibliography}
\end{document}